# Quantum Criticality and Universal Scaling of a Quantum Antiferromagnet

[1+]Bella Lake, [2,3*]D. Alan Tennant, [2]Chris D. Frost and [1]Stephen E. Nagler

[1] Condensed Matter Sciences Division, Oak Ridge National Laboratory, Oak Ridge, Tennessee 37831-6393, USA.
[2] ISIS Facility, Rutherford Appleton Laboratory, Chilton, Didcot, Oxfordshire OX11 0QX, U.K.
[3] Clarendon Laboratory, Parks Road, Oxford OX1 3PU, U.K.

**Quantum effects dominate the behaviour of many diverse materials. Of particular current interest are those systems in the vicinity of a quantum critical point (QCP). Their physical properties are predicted to reflect those of the nearby QCP with universal features independent of the microscopic details. The prototypical QCP is the Luttinger liquid (LL) which is of relevance to many quasi-one-dimensional materials. The magnetic material $KCuF_3$ realizes an array of weakly-coupled spin chains (or LLs) and thus lies close to but not exactly at the Luttinger liquid quantum critical point. By using inelastic neutron scattering we have collected a complete data set of the magnetic correlations of $KCuF_3$ as a function of momentum, energy, and temperature. The LL description is found to be valid over an extensive range of these parameters, and departures from this behaviour at high and low energies and temperatures have been identified and explained.**

The concept of a quantum critical point[1,2] - a zero temperature phase transition between quantum ground states – provides an attractive approach for describing regions dominated by quantum mechanics in condensed matter. In its vicinity universal behaviors are predicted that are independent of the microscopic details of the system. An example of this is universal energy/temperature ($E/T$) scaling of the collective response of the system – a phenomenon that has received heightened attention due to its observation in both cuprate high-temperature superconductors[3,4,5] and unconventional metals[6,7,8]. By exploiting the powerful mathematics of scaling, the aim is to set up systematic expansions of the physical properties about these special points thus explaining the quantum-dominated behaviour found in a diverse range of systems. In quasi-one-dimensional quantum magnets experimental and theoretical techniques[9] are now sufficiently advanced to study these ideas and methods in depth. Here we consider the generic problem of an array of one-dimensional systems weakly coupled together into three dimensions.

Quantum critical states in one-dimension (1D) are thought to apply to systems as diverse as carbon nanotubes, stripes in cuprate high-temperature superconductors, and spin chains[10,11]. The prototypical example is the spin-1/2 ($S$=1/2) Heisenberg antiferromagnetic chain (HAFC) which maps onto the Luttinger liquid QCP. Here the magnetic ions possess spin angular momentum of 1/2, and interact with their nearest neighbours via antiferromagnetic Heisenberg exchange couplings in only one crystallographic direction. This system fails to develop long-range Néel ordering (where

---

[+] Present address - Clarendon Laboratory, Parks Road, Oxford OX1 3PU, U.K.
[*] Present address - School of Physics & Astronomy, North Haugh, St Andrews, Fife KY16 9SS, U.K.

neighbouring spins point anti-parallel to each other) even at the lowest temperatures and has exceptional dynamical properties. The basic excitations are spinons[12], which are topological excitations that can be visualized as twists of π in the spin order. Spinons are fractional particles which possess spin values of $S=1/2$ and since in quantum mechanics changes in angular momentum are restricted to integer units, only an even number of spinons can be created[13].

The multi-spinon excitation spectrum forms a continuum over an extended region of energy ($E$) and momentum or wavevector ($q$) space. The excitation spectrum is reflected in the dynamical structure factor probed by neutron scattering. At $T=0$ K this is given approximately by the Müller Ansatz[14]

$$S(q,E) = \frac{A_M \Theta(E - E_l(q))\Theta(E_u(q) - E)}{\sqrt{E^2 - E_l^2(q)}} \tag{1}$$

where $\Theta$ is the Heaviside step function and $A_M$ is a constant. The quantities $E_l$ and $E_u$ are the lower and upper energy boundaries of the continuum and are sinusoidal functions of wavevector given by

$$E_l(q) = \frac{\pi J}{2}|\sin(qc)| \quad \text{and} \quad E_u(q) = \pi J \left|\sin\left(\frac{qc}{2}\right)\right| \tag{2}$$

where c is the lattice parameter in the chain direction and $J$ is the antiferromagnetic exchange constant coupling nearest neighbours along the chain. The two-spinon contribution that dominates $S(q,E)$ has been calculated exactly by Karbach et al.[15] and found to extend between the same upper and lower boundaries, with an identical singularity at $E_l(q)$ and a smooth cutoff at $E_u(q)$. In order to develop the model for finite temperatures, field-theory techniques are used where the discrete lattice is approximated by a continuous medium. Schulz predicts that at the antiferromagnetic zone centre (AFZC) $q_{AFZC}=\pi/c$, the dynamical structure factor is given by

$$S(\pi,E) = \frac{e^{E/kT}}{e^{E/kT}-1}\frac{A}{T}\text{Im}\left[\rho\left(\frac{E}{4\pi T}\right)^2\right] \tag{3}$$

where $\rho(x)=\Gamma(1/4-ix)/\Gamma(3/4-ix)$ and $A$ is a constant[16]. It is clear from this equation that the structure factor multiplied by temperature depends only on the dimensionless ratio of $E$ to $T$ rather than on these quantities separately and therefore obeys universal scaling. The ideal $S=1/2$ HAFC is rarely if ever found in the real world since inevitably any Hamiltonian contains extra terms which have the potential to change the physical state; for example even the smallest amount of interchain coupling alters the ground state and gives rise to Néel order (although with reduced ordered spin moment). The first problem we address in this paper is the experimental confirmation of the scaling specific to a 1D LL in a real material which is close to but not at this QCP.

Dimensionality has important consequences for antiferromagnetism. For systems with nearest neighbour exchange interactions this is evident from simple topological considerations since the number of magnetic ions with which each spin interacts directly is an important factor in stabilizing long-range order. The ideal 1D $S=1/2$ antiferromagnet, with only two nearest-neighbours for each spin, fails to develop Néel order. Conversely, in the simple cubic three-dimensional (3D), $S=1/2$ antiferromagnet where there are six nearest neighbours, mean-field effects lead to long-range order and the Néel state is often an adequate approximation to the true ground state. The excitations in the ordered state are spin-waves (Goldstone modes) which possess spin values of 1 and follow a well-defined trajectory in energy and wavevector[17,18] in complete contrast to the multi-spinon continuum. Quantum fluctuations are largely suppressed in this system and a semi-classical description in terms of the 3D non-linear sigma model (NLσM)[19] is appropriate. The second problem that we address in this paper is the evolution of the magnetic correlation function with momentum, energy, and temperature under the influence of interchain coupling.

To study these problems we investigate the magnetic excitation spectrum of the quasi-1D, $S=1/2$, Heisenberg antiferromagnet $KCuF_3$. The magnetic $Cu^{2+}$ ions carry $S=1/2$ and are coupled into chains by a strong antiferromagnetic exchange interaction $J=34$ meV, there is also a weak ferromagnetic exchange interaction $J_\perp=-1.6$ meV which acts to couple these chains together[18,20]. The magnetic Hamiltonian is

$$H = J\sum_{n,r} \vec{S}_{n,r} \cdot \vec{S}_{n+1,r} + J_\perp \sum_{n,r,\delta} \vec{S}_{n,r} \cdot \vec{S}_{n,r+\delta} \qquad (4)$$

where $n$ labels the sites along the chain, $r$ represents a lattice vector lying in the plane perpendicular to the chains and $\delta$ is summed over the four nearest neighbours in this plane. Magnetic ordering occurs below the Néel Temperature of $T_N=39$ K with a saturated moment of 0.5 $\mu_B/Cu^{2+}$ at $T=4$ K[20]. The suppressed ordering temperature and moment reduction of 50% (from Néel ordering) show that despite the interchain coupling, $KCuF_3$ exhibits strong quantum fluctuations and is in the proximity of a QCP.

To measure the magnetic excitations in $KCuF_3$ we used the experimental technique of inelastic neutron scattering, for which the cross-section is directly proportional to the dynamical structure factor and can be used to probe it as a function of wavevector (1/distance) and energy (1/time) transfer. Figure 1 shows data collected over the full energy and wavevector range of the excitations at 6K, 50K, 150K and 300K (data was also collected at 75K, 100K, 200K). The data is presented as a function of energy and wavevector parallel to the chains and is integrated over the two-dimensional Brillouin zone perpendicular to the chains. Figure 2b shows a detailed measurement at $T=11$K of the low energy spectrum close to the antiferromagnetic zone centre. In all cases corrections have been made for background and the magnetic form factor of copper.

The data show that distinct regimes of behaviour occur in different regions of energy and temperature space. The extended scattering at high energies in Fig. 1a-c is similar to the excitation continuum of the 1D LL, indeed the scattering is bounded by the black dashed lines marking the theoretical upper and lower boundaries of the spinon continuum of a 1D $S=1/2$ HAFC in its ground state given by the Muller Ansatz (equation (1)). While this result might be expected at temperatures above the Néel temperature it is clear that the continuum scattering is also present below $T_N$ where long-range magnetic order develops (Fig. 1a). However detailed measurements made at low energies below $T_N$, reveal highly dispersive V-shaped scattering (Fig. 2b) typical of the spin-waves of the 3D NLσM. Such coexistence of 1D LL and 3D NLσM behavior has been observed previously in $KCuF_3$[21,22] and is a feature of the quasi-1D $S=1/2$ HAFC; similar observations have been made in copper benzoate[23] and $BaCu_2Si_2O_7$[24].

To test whether the 1D LL model is appropriate for $KCuF_3$ at high energies, the data was compared to the field theory result given by equation (3). The data was summed over the AFZC parallel to the chain direction ($0.48<qc/2\pi<0.52$) and integrated over the entire Brillouin zone perpendicular to the chain. This is plotted as a function of energy for the $T$=50K data in the inset of Fig. 3 and the solid line represents Equation (3). Data lies on the theoretical curve for 26<$E$<80 meV proving that the 1D LL description is valid over this energy range, however the data fails to follow this theory at higher and lower energies. The data sets for other temperatures were treated in a similar manner and all of them except the 300K data were found to follow the 1D LL expression over a large range of energies. In each case the lower limit of this range was chosen as the energy below which two successive data points were more than 1.5 standard deviation above the fitted line. The main part of Fig. 3 shows the combined data for which the 1D theory gives an accurate description, multiplied by temperature and plotted as a function of the universal scaling parameter $E/T$, again the solid line is equation (3). What is remarkable is that although the data sets range in temperature from 6K to 200K they all lie on the same solid line with the same constant of proportionality A; i.e they depend only on the ratio of energy to temperature but not on these quantities independently. This result shows that the theoretical concept of the 1D LL strongly influences the physics of real magnetic systems which lie close to but not at this QCP due to a small amount of interchain coupling. It also clearly proves experimentally that the magnetic excitation spectrum of a material in this regime obeys universal E/T scaling over an extensive range of energies and temperatures.

    The analysis described in the previous paragraph shows that by testing the magnetic excitation spectrum for scaling it is possible to identify not only whether a magnetic system behaves like a 1D LL but also the energy and temperature ranges over which the QCP dominates. Thus this technique can be used to identify the different regimes of behaviour for the correlations of a quantum magnet and construct a "magnetic crossover diagram" showing where the different descriptions are dominant. Figure 4 shows this diagram for $KCuF_3$. Besides the 1D LL regime it has also been possible to identify a 3D NLσM regime which occurs at low energies and temperatures and is bounded by an upper temperature of $T_N$ above which long-range magnetic order is lost. The existence of this phase was established by measuring the sharpness of the excitations. δ-function modes, observed in neutron scattering as resolution-limited peaks, are a signature of spin-wave excitations and for the $T$=6K data they are found at energies of 11meV and below. Above this, extra scattering is observed at the AFZC and the spin-wave model can no longer quantitatively account for the data. The scattering can be seen directly in fig 2b where, as energy transfer is increased at the AFZC, the intensity first decreases as the spin-wave branches disperse apart, however above 11 meV the contours indicate an increase in the scattering intensity.

    While excitations with energies and temperatures that place them just outside the 3D NLσM regime cannot be explained by a spin-wave model they also do not show the scaling behaviour of the 1D LL (equation (3)). We denote this intermediate state between the 3D NLσM and 1D LL regimes, the crossover region (fig. 4) since physically it corresponds to the onset of deconfinement where the $S$=1 spin-waves begin to fractionalize into pairs of $S$=1/2 spinons. This regime is characterised at low temperatures by the presence of a novel longitudinal mode - an excitation whose existence has only recently been established theoretically and experimentally[25,26,27,28]. The mode appears for $T$<$T_N$ lying at an energy of $E$=15 meV (see Fig. 2b) and is broadened in energy with a full-width at half-maximum intensity of 5 meV[25], indicating a lifetime reduced by decay

processes. The longitudinal mode is a crossover phenomenon existing in the presence of long-range magnetic order but where the size of the ordered moment is significantly less than the full spin moment. Long-range order is required because longitudinal excitations are defined as oscillations in the length of this ordered moment, however if the full spin moment of each magnetic ion were to point along the ordering direction then these oscillations would clearly be restricted and their intensity negligible.

The ranges in energy and temperature occupied by the various regimes in $KCuF_3$ are very significant. Broadly speaking the 1D LL description is valid at high energies and temperatures where the effects of interchain coupling become negligible whereas the 3D NLσM description is valid at low energies and temperatures where the interchain interaction energy is similar to the thermal and excitation energies. The 3D NLσM regime exists for $T<T_N$ where long-range order occurs, and $E\leq11$ meV where no longitudinal mode or spinon continuum scattering can be observed. Significantly this energy is similar to $M=11$meV the maximum (zone boundary) energy of the spin-waves perpendicular to the chain direction due to interchain coupling. The 1D LL regime occurs for $26<E<80$ meV (at $T=6$K). The lower limit of this regime is identified by the onset of scaling and is similar to $2M=22$ meV which is the lowest energy that a spinon-pair must have if both spinons are to have energies greater than the perpendicular zone boundary energy.

The upper edge of the 1D LL regime is also interesting and has its roots in the discrete nature of the crystal lattice. The 1D LL QCP maps directly onto the field theory expression (equation (3)). This equation however provides only an approximation to the behaviour of a S=1/2 HAFC where the discrete lattice has been replaced by a continuous medium so that the single-spinon excitation energies vary linearly with wavevector rather than sinusoidally as is the case for a lattice. This contrasts with the ground state expression (equation (1)) which takes the lattice into account. The continuum assumption is adequate for wavevectors around the AFZC where the length scale of the excitations is much greater than the lattice spacing, but will clearly start to deviate away from the AFZC. Simulations suggest that the deviation is significant for wavevectors $|q-q_{AFZC}| >0.15\times2\pi/c$, for which the single-spinon energy is ~40meV. The implication is that equation (3) can be expected to fail for energies greater that the corresponding spinon-pair energy of ~80meV. It follows that 1D LL scaling is not obeyed above this energy. This is manifest in the measured data as a reduction in spectral weight at the highest energies compared to the predictions of equation (3). Figs 1a-c show a drop in intensity (contour line) around $E=75-80$meV for all data sets except the $T=300$K data. This is a gradual crossover and the upper limit of the 1D LL regime was set at ~80meV which is the average energy of all the data sets at which two consecutive points fall below the scaling expression by more than 1.5 standard deviations. Finally as mentioned before, the scaling expression fails to account for any of the data collected at $T=300$K. The explanation for this is that thermal fluctuations dominate over quantum fluctuations at high temperatures and the magnetic response becomes paramagnetic. The Curie-Weiss temperature can be used as a rough guide to the onset of paramagnetism and for $KCuF_3$ it is calculated to be $T_{Curie-Weiss}=216$K[29] using the experimental values for the exchange constants. At temperatures greater than this, thermal fluctuations are expected to destroy the 1D LL behaviour.

Our findings are consistent with QCP theory that has been modified to take account of interchain and other couplings[27]. The procedure is to map each magnetic chain onto a quantum 'string'[9] and use the powerful mathematics of conformal field theory to calculate quantities such as the collective response (Equation (3)). This method can take into account all the relevant couplings that displace the $S$=1/2 HAFC away from the quantum critical state. Essler, Tsvelik and Delfino (ETD)[27] have considered the coupling between chains in $KCuF_3$ and show that this particular operator acts to drive the system away from criticality, giving rise to magnetic order. They are able to calculate the response functions at $T$=0 K using the method of form factors[30] and they find that in the ordered phase, the effect of ordered neighbouring chains is to confine spinons on long time and distance scales into 3D spin-waves below 11.5 meV, just as we find experimentally. Physically this is because as spinons separate they leave overturned regions of spins that are mismatched with the order, this is expressible as a potential that confines spinons which is proportional to their separation distance. ETD also predict a longitudinal mode, a novel type of bound mode of spinons, at 17 meV which we indeed observe at the slightly lower energy of 15 meV in the crossover regime[25]. For energies above 22meV the response is expected to resemble the uncoupled 1D LL since on these short time-scales the spinons behave as if nearly free, which again corresponds to our findings. In fact expansions around the 1D LL QCP account also to high accuracy for the size of the ordered spin moment and Néel temperature[28,31].

To conclude we have made the most detailed survey of the magnetic response of a quasi-1D, $S$=1/2, Heisenberg antiferromagnet as a function of energy, temperature and wavevector. The data have been used to establish that the 1D LL description is valid for this system over an extensive range of temperatures and energies, and that the magnetic correlations obey the universal E/T scaling predicted for this QCP. Indeed by testing for scaling we have been able to construct the first magnetic crossover diagram of such a system, where the limits of the 1D LL regime are determined by the energies and temperatures where scaling breaks down, and the neighbouring phases have been identified and understood. Our measurements give firm experimental support to the QCP concept and show that it influences the behaviour of real materials that lie close to the QCP.

**Methods**

The neutron scattering data shown in figure 1 and figure 2a was collected using the MAPS neutron scattering spectrometer at the ISIS neutron spallation source in the Rutherford Appleton Laboratory, U.K. MAPS is a new state-of-the-art spectrometer with unprecendented detector coverage allowing large expanses of energy and wavevector space to be measured simultaneously to give the complete excitation spectrum. The crystal was aligned with its chains perpendicular to the incident neutron beam. A Fermi chopper was phased to select neutrons with an incident energy of 150 meV and was rotated at 500 Hz to provide an energy resolution of 5 meV. The sample was cooled in a closed cycle cryostat and was measured at a variety of temperatures from 6K to 300K, each temperature was collected for about a day (~3000 μAmphr). The data shown in figure 2b was collected at $T$=11K using the HB1 triple-axis spectrometer at the High-Flux Isotope Reactor in Oak Ridge National Laboratory, U.S.A.[25]. The crystal was aligned so that chains lay in the horizontal scattering plane. The incident and final energies were selected using a monochromator and an analyser made of Pyrolytic Graphite (PG). The final energy was fixed at 13.5 meV and a PG filter was placed after the sample to remove higher order neutrons from the beam. A series of collimators with values 48´-40´-40´-120´ were used between the source and detector to provide an energy resolution of 1.3 meV and a wavevector resolution of 0.057 Å$^{-1}$ parallel to the chains at an energy transfer of 16 meV.

**Acknowledgements**
We thank J.S. Caux, R. Coldea, F.H.L. Essler and A.M Tsvelik for helpful discussions and G. Shirane for the loan of the crystal. ORNL is operated by UT-Battelle LLC., under contract no. DE-AC05-00OR22725 with the U.S. Department of Energy.

**Competing interests statement**
The authors declare that they have no competing financial interests.

Correspondence and requests for materials should be addressed to B.L.
(e-mail: bella.lake@physics.ox.ac.uk).

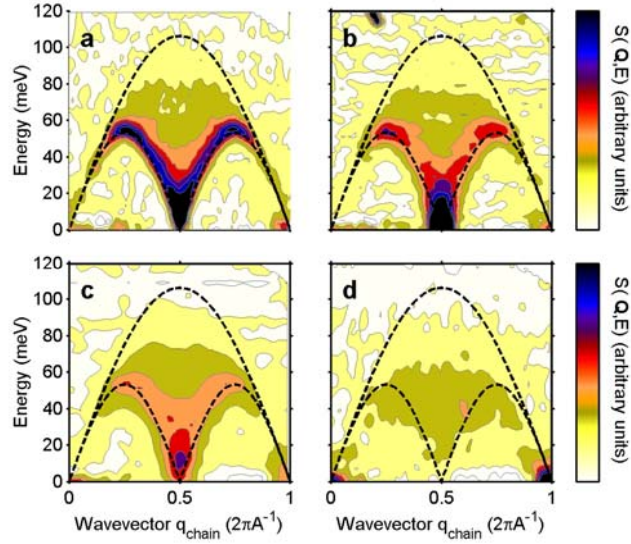

**Figure 1**
**Inelastic neutron scattering data for $KCuF_3$.** The data is plotted as a function of $E$ and $q$ parallel to the chains for the temperatures **a,** $T$=6K, **b,** $T$=50K, **c,** $T$=150K and **d,** $T$=300K. The colours indicate the size of the neutron scattering cross-section $S(q,E)$ and the superimposed black dashed lines indicate the region where the multi-spinon continuum is predicted at $T$=0K by the Muller Ansatz (equation (1)). The data was collected using the MAPS time-of-flight spectrometer at ISIS, Rutherford Appleton Laboratory, U.K.

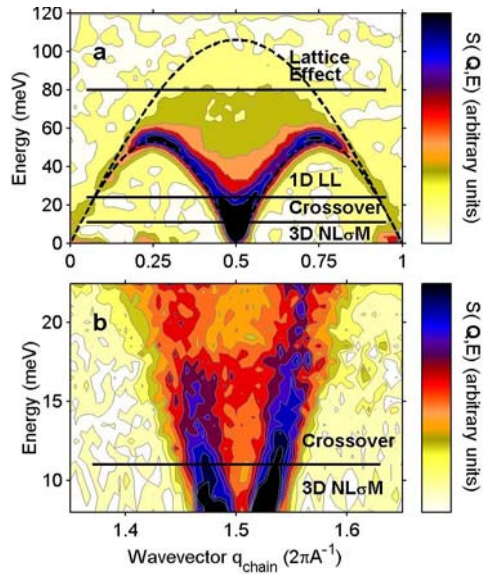

**Figure 2**
**Inelastic neutron scattering data for $KCuF_3$ measured well below $T_N$ and plotted as a function of energy and wavevector parallel to the chain direction.** Again the colours indicate the size of the neutron scattering cross-section $S(q,E)$ and the superimposed black dashed lines indicate the region where the multi-spinon continuum is predicted at $T$=0K by the Muller Ansatz (equation (1)). The solid horizontal black lines are used to mark the energy ranges of the different regimes that exist at low temperatures and these are labeled on the diagram. **a,** shows data collected at $T$=6K using the MAPS time-of-flight spectrometer at ISIS, Rutherford Appleton Laboratory, U.K. **b,** shows data collected at $T$<11K using the HB1 triple axis spectrometer at HFIR, Oak Ridge National Laboratory, U.S.A. The intensity scales on the two data sets are different.

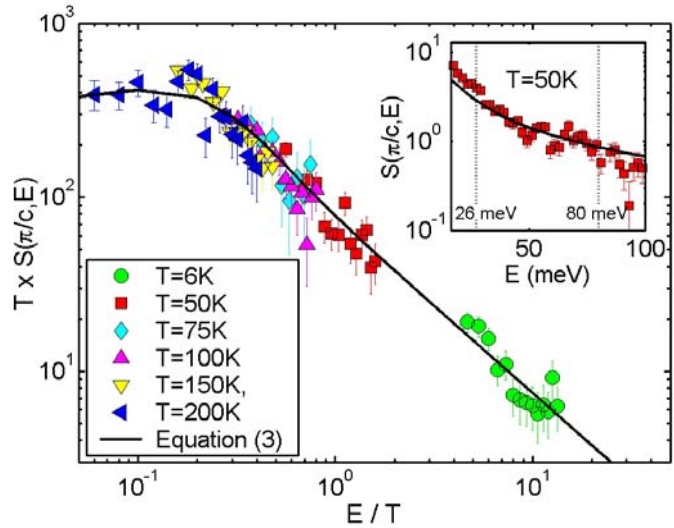

**Figure 3**
**Universal energy/temperature scaling in KCuF$_3$.** Inset figure: The inset shows the corrected data at $T$=50K plotted as a function of energy. The data is integrated over the entire Brillouin zone perpendicular to the chain direction and summed over a narrow range at the AFZC parallel to the chain direction (0.48<$q$c/2π<0.52). The solid line through the data is the field theory expression for an ideal 1D $S$=1/2 HAFC given by equation (3) which maps onto the 1D LL. The data follows this line for energies in the range 26<$E$<80 meV (bounded by the vertical dashed lines) indicating that the 1D LL description is valid for KCuF$_3$ at these energies. Main figure: The data for all the other temperatures (except 300K) was treated in a similar manner and was also found to follow equation (3) over a defined energy range. The combined data showing 1D LL behaviour is multiplied by temperature and plotted as a function of the universal parameter $E/T$ in the main part of the figure. Again the solid line through the data is equation (3).

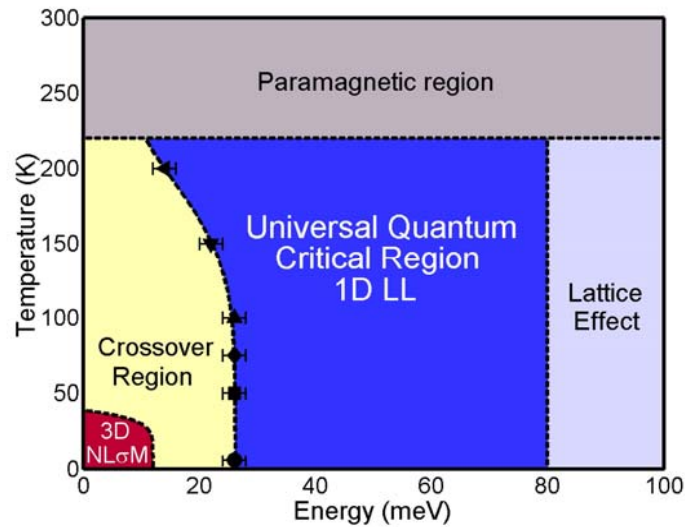

**Figure 4**
**Magnetic crossover diagram showing the different physical regimes of KCuF$_3$ as a function of energy and temperature.** The methods described in the caption of figure 3 and the text, have been used to identify the different magnetic regimes and the regions in energy and temperature that they occupy. In particular the solid black symbols represents the lower energy limit of the 1D LL regime identified by the breakdown of scaling (see the inset of fig 3 which illustrates this for the $T$=50 K dataset). The regimes are labeled on the diagram and the dashed lines mark the places where one phase crosses over into another.